\documentclass[12pt,a4paper]{article}
\usepackage{axodraw}
\usepackage{graphicx}
\usepackage{amsmath}
\usepackage{amsfonts}
\usepackage{amssymb}
\newcommand{\vs}{\vspace{-0.25cm}}

\newcommand{\fmd}{\,\mathrm{fm}^{-3}}

\newcommand{\dif}{\mathrm{d}}

\begin{document}

\begin{center}

{\Large
\textbf{Chiral approach to nuclear matter:\\ Role of explicit short-range 
NN-terms}\footnote{Work supported in part by BMBF, GSI and DFG.} }

\bigskip

{\large S. Fritsch$\,^{a,b}$ and N. Kaiser$\,^a$}\\

\bigskip

{\small $^a$\,Physik Department T39, Technische Universit\"{a}t M\"{u}nchen,
D-85747 Garching, Germany\\
\smallskip

$^b$\,ECT$^*$, I-38050 Villazzano (Trento), Italy\\

\smallskip

{\it email: nkaiser@physik.tu-muenchen.de}}

\end{center}

\bigskip

\begin{abstract}
We extend a recent chiral approach to nuclear matter by including the most
general (momentum-independent) NN-contact interaction. Iterating this 
two-parameter contact-vertex with itself and with one-pion exchange the
emerging energy per particle exhausts all terms possible up-to-and-including 
fourth order in the small momentum expansion. Two (isospin-dependent) cut-offs
$\Lambda_{0,1}$ are introduced to regularize the (linear) divergences  of some 
three-loop in-medium diagrams. The equation of state of pure neutron matter,
$\bar E_n(k_n)$, can be reproduced very well up to quite high neutron densities
of $\rho_n=0.5\fmd$ by adjusting the strength of a repulsive $nn$-contact 
interaction. Binding and saturation of isospin-symmetric nuclear matter is a 
generic feature of our perturbative calculation. Fixing the maximum binding 
energy per particle to $-\bar E(k_{f0})= 15.3\,$MeV we find that any possible 
equilibrium density $\rho_0$ lies below $\rho_0^{\rm max}=0.191\fmd$. The 
additional constraint from the neutron matter equation of state leads however
to a somewhat too low saturation density of $\rho_0 =0.134 \fmd$. We also 
investigate the effects of the NN-contact interaction on the complex 
single-particle potential $U(p,k_f)+i\,W(p,k_f)$. We find that the effective 
nucleon mass at the Fermi-surface is bounded from below by $M^*(k_{f0}) \geq 
1.4 M$. This property keeps the critical temperature of the liquid-gas phase 
transition at somewhat too high values $T_c \geq 21\,$MeV. The downward bending
of the asymmetry energy $A(k_f)$ above nuclear matter saturation density is a 
generic feature of the approximation to fourth order. We furthermore 
investigate the effects of the NN-contact interaction on the $(\vec \nabla
\rho)^2$-term in the nuclear energy density functional ${\cal E}[\rho,\tau]$.  
Altogether, there is within this complete fourth-order calculation no "magic" 
set of adjustable short-range parameters with which one could reproduce
simultaneously and accurately all semi-empirical properties of nuclear
matter. In particular, the conditions for a good neutron matter equation of
state and for good single-particle properties are mutually exclusive.   
\end{abstract}

\medskip


\medskip

\section{Introduction}
The present status of the nuclear matter problem is that a quantitatively
successful description can be achieved, using advanced many-body techniques, 
in a non-relativistic framework when invoking an adjustable three-nucleon force
\cite{akmal}. Alternative relativistic mean-field approaches, including either
non-linear terms with adjustable parameters or explicitly density-dependent
point-couplings, are also widely used for the calculation of nuclear matter
properties and of finite nuclei \cite{ring,walecka,typel,lenske}. At a more 
basic level, the Dirac-Brueckner method of ref.\cite{brockmann} solves a
relativistically improved Bethe-Goldstone equation with one-boson exchange
NN-interactions.    

In recent years a novel approach to the nuclear matter problem based on
effective field theory (in particular chiral perturbation theory) has emerged
\cite{lutz,nucmat,pot}. The key element there is a separation of long- and 
short-distance dynamics and an ordering scheme in powers of small momenta. At 
nuclear matter saturation density the Fermi-momentum $k_{f0}$ and the pion mass
$m_\pi$ are comparable scales ($k_{f0}\simeq 2m_\pi$), and therefore pions
must be included as explicit degrees of freedom in the description of the 
nuclear many-body dynamics. The contributions to the energy per particle $\bar
E(k_f)$ of isospin-symmetric nuclear matter as they originate from chiral
pion-nucleon dynamics have been computed up to three-loop order in
refs.\cite{lutz,nucmat}. Both calculations include the $1\pi$-exchange
Fock-diagram and the iterated $1\pi$-exchange Hartree- and Fock-diagrams. In
ref.\cite{nucmat} irreducible $2\pi$-exchange has also been taken into account
and a momentum cut-off $\Lambda$ has been used to regularize the few divergent
parts associated with chiral $2\pi$-exchange. The resulting cut-off dependent
contribution ($\bar E(k_f)\sim\Lambda k_f^3$, see eq.(15) in ref.\cite{nucmat})
is completely equivalent to that of a zero-range NN-contact interaction. At
that point the (earlier) work of Lutz et al. \cite{lutz} follows a different 
strategy. Two zero-range NN-contact interactions (acting in the $^3S_1$ and 
$^1S_0$ NN-states) proportional to the dimensionless parameters $g_0+g_A^2/4$ 
and $g_1+g_A^2/4$ are explicitly introduced (see eq.(4) in ref.\cite{lutz}). 
The components proportional to $g_A^2/4$ cancel (order by order) zero-range 
contributions generated by one-pion exchange. The other components 
proportional to $g_0$ and $g_1$ are understood to subsume all non-perturbative 
short-range NN-dynamics relevant at densities around nuclear matter saturation
density $\rho_0$.  

Despite their differences in the treatment of the effective short-range 
NN-dynamics both chiral approaches \cite{lutz,nucmat} are able to reproduce
correctly several empirical nuclear matter properties (saturation density 
$\rho_0$, binding energy per particle $-\bar E(k_{f0})$ and compressibility 
$K= k_{f0}^2 \bar E''(k_{f0})$) by fine-tuning only one parameter, either the 
coupling $g_0+g_1 \simeq 3.23$ or the cut-off $\Lambda \simeq 0.65\,$GeV.  
The fine-tuning of a momentum cut-off $\Lambda$ as done in ref.\cite{nucmat}
introduces some degree of non-uniqueness that lies outside chiral perturbation
theory. One should, however, keep in mind that fine-tuning of some parameter(s)
is involved in most calculations of nuclear matter.   

Although the conceptual treatment of the short-range NN-interaction by Lutz et
al. \cite{lutz} may be very appealing at first sight it suffers (after closer
inspection) from severe phenomenological problems. As shown recently in
ref.\cite{lutzcontra} the single-particle properties come out completely
unrealistic in this scheme. The potential depth of $U(0,k_{f0}) = -20\,$MeV is
by far too weakly attractive while on the other hand the imaginary
single-particle potential $W(0,k_{f0})= 51\,$MeV is too large. Most seriously,
the total single-particle energy $T_{\rm kin}(p)+U(p,k_{f0})$ does not rise
monotonically with the nucleon momentum $p$, thus implying a negative effective
nucleon mass at the Fermi-surface $p= k_{f0}$ (see Fig.\,3 in
ref.\cite{lutzcontra}). In such an abnormal situation there exist occupied
nucleon-states in the Fermi-sea with total energy higher than the Fermi-energy,
indicating an instability of the system. As a matter of fact the overly strong
momentum dependence of the nuclear mean-field $U(p,k_f)$ in the scheme of Lutz
et al. \cite{lutz} (leading to $\partial U/\partial p<-p/M$) originates from
those diagrams in which the (strong and) attractive NN-contact interaction
(proportional to $g_0+g_1+g_A^2/2$) is further iterated. The NN-contact
interaction necessarily has to be very strong in this scheme since it
effectively also has to account for the attraction generated dynamically by
iterated $1\pi$-exchange in form of a (regularization dependent) linear
divergence $\int_0^\infty{\dif}l\,1$. In contrast to that much better
single-particle properties have been obtained in the chiral approach of
ref.\cite{pot} employing cut-off regularization and not introducing any
explicit short-range NN-terms. For example, the resulting potential depth of
$U(0,k_{f0}) =-53\,$MeV \cite{pot} is in good agreement with the depth of the
empirical nuclear shell model potential \cite{bohr} or the optical model
potential \cite{hodgson}. Furthermore, when extended to finite temperatures
this approach reproduces the first-order liquid-gas phase transition of
isospin-symmetric nuclear matter, however, with a somewhat too high value of
the critical temperature $T_c =25\,$MeV \cite{liquidgas}. The reason for that
is a still too strong momentum dependence of the real single-particle potential
$U(p,k_{f0})$ near the Fermi-surface $p=k_{f0}$ \cite{pot}.   

The purpose of the present paper is to investigate in detail the role of 
short-range NN-terms in addition to pion-exchange dynamics for a variety of
nuclear matter observables. Our primary interest is to find out whether such
adjustable NN-contact terms (of moderate strength) can cure certain
shortcomings of previous calculations \cite{nucmat,pot} which have included
only the pion-exchange contributions to three-loop order. From the point of
view of the expansion in powers of small momenta ($k_f$ and $m_\pi$) the
combination of pion-exchange and zero-range NN-contact interaction exhausts 
actually all terms possible up-to-and-including fourth order. Alternatively, 
the interacting part of the energy per particle can be understood (within this
approximation to order ${\cal O}(k_f^4))$ as to result from the Hartree and
Fock contributions of a nucleon-nucleon  T-matrix of the form:
\begin{equation} {\cal T}_{NN}= {g_A^2 \over 4f_\pi^2} \bigg[ {\vec
\sigma_1 \cdot \vec q\,\, \vec \sigma_2 \cdot \vec q \over m_\pi^2 +\vec q\,^2}
\,\vec \tau_1 \cdot \vec\tau_2 +{\gamma_0+3\gamma_1\over 2} +{\gamma_1-\gamma_0
\over 2}\,\vec \tau_1 \cdot \vec \tau_2 \bigg] \,,  \end{equation}
which is evaluated in first and second order perturbation theory (i.e. it is 
also iterated once with itself). Note that additional spin-spin terms 
proportional to $\vec \sigma_1 \cdot \vec \sigma_2$ are redundant in the 
T-matrix eq.(1) as a consequence of the Pauli exclusion principle. The two 
coefficients, $\gamma_0$ and $\gamma_1$, parameterize the strength of the 
contact interaction in NN-states with total isospin $I=0,1$. In order to have 
immediately a measure of the (relative) magnitudes involved we note that the 
repulsive contact piece which survives from one-pion exchange in the chiral
limit $m_\pi=0$ can be mapped onto $\gamma_0^{(1\pi)} =\gamma_1^{(1\pi)}=
-1/2$. As usual, $f_\pi= 92.4\,$MeV denotes the pion decay constant and $m_\pi
=135\,$MeV stands for the (neutral) pion mass. The value $g_A=1.3$ of the
nucleon axial-vector coupling constant is linked via the Goldberger-Treiman
relation to a $\pi NN$-coupling constant of $g_{\pi N} = g_A M/f_\pi= 13.2$
\cite{pavan}. The iteration of the T-matrix eq.(1) to second order generates
besides finite terms also the (primitive) linearly divergent loop-integral
$\int_0^\infty {\dif}l\,1$. As done in ref.\cite{nucmat} it will be regularized
by a momentum cut-off $\Lambda_{0,1}$ which we allow in this work to depend on
the total isospin $I=0,1$ of the two-nucleon system.\footnote{The weighting
factors of $\Lambda_0$ and $\Lambda_1$ are readily found by decomposing the
isospin factor of a particular diagram into the projection operators on total
isospin $I=0,1$: ${\cal P}_0 = (1-\vec \tau_1 \cdot \vec \tau_2)/4$ and ${\cal
P}_1 = (3+\vec \tau_1 \cdot\vec\tau_2)/4$.} Such a doubling procedure is
physically meaningful since cut-off dependent contributions are fully
equivalent to contributions from (momentum independent) NN-contact vertices. On
the practical side four free parameters, $\gamma_{0,1}$ and $\Lambda_{0,1}$
give us more flexibility in adjusting nuclear matter properties (within the
present perturbative scheme). Of course, the two cut-offs $\Lambda_{0,1}$ 
should not differ too much from each other and they should also take on 
physically reasonable values.   
 
\begin{figure}
\begin{center}
\includegraphics[scale=1.5,clip]{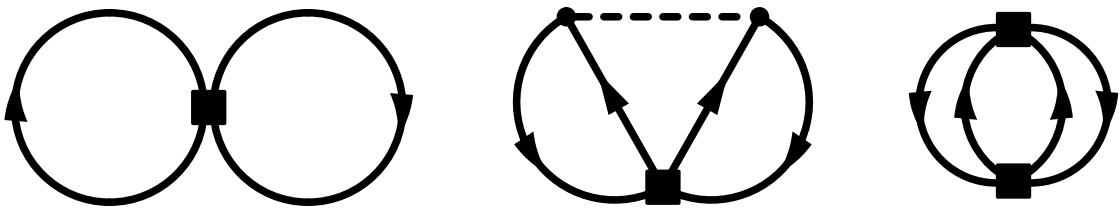}
\end{center}\vspace{-0.2cm}
\caption{Additional in-medium diagrams generated by the NN-contact interaction.
The filled square vertex symbolizes this zero-range NN-contact interaction 
proportional to $\gamma_0$ and $\gamma_1$. The indices $I=0,1$ label the total 
isospin of the NN-system.}
\end{figure}
\section{Neutron matter equation of state}
We start the discussion with the equation of state of pure neutron matter. For
this system the total isospin of two nucleons is restricted to $I=1$ and 
therefore only the strength parameter $\gamma_1$ (see eq.(1)) and the isospin-1
cut-off $\Lambda_1$ can come into play. We first write down the contributions 
to the energy per neutron $\bar E_n(k_n)$ generated by the zero-range $nn
$-contact interaction proportional to $\gamma_1$. The left diagram in Fig.\,1 
obviously leads to a contribution linear in the neutron density $\rho_n = k_n^3
/3\pi^2$. Of the same structure are all the cut-off dependent terms from 
iterated diagrams with two medium insertions\footnote{This is a technical 
notation for the difference between the in-medium and vacuum nucleon 
propagator. For further details, see sect. 2 in ref.\cite{nucmat}.}. We combine
them by introducing the dimensionless parameter $\Gamma_1$: 
\begin{equation} \bar E_n(k_n) = - {\Gamma_1 g_A^2 k_n^3 \over 6(2\pi f_\pi)^2}
\,, \qquad \Gamma_1 = \gamma_1 +{g_A^2 M \Lambda_1 \over (4\pi f_\pi)^2 }
(2\gamma_1 -1)^2\,.      \end{equation}
The second diagram in Fig.\,1 with two medium insertions on parallel 
nucleon lines gives rise to a finite (i.e. cut-off independent) contribution
of the form:
\begin{equation} \bar E_n(k_n) = {\gamma_1 g_A^4 M m_\pi^4 \over 5(8\pi)^3 
f_\pi^4} \bigg[ 22u -{1\over u}-4(5+4u^2) \arctan 2u +\bigg( {1\over 4u^3}
+{5\over u} \bigg) \ln(1+4u^2) \bigg] \,,\end{equation}
with $u=k_n/m_\pi$ the ratio of the neutron Fermi-momentum $k_n$ and the pion
mass $m_\pi$. These two quantities are treated as small momentum/mass scales 
in our perturbative calculation. The second and third diagram in Fig.\,1 with 
three medium insertions represent Pauli-blocking effects. The corresponding
contribution to the energy per neutron reads:
\begin{eqnarray} \bar E_n(k_n)&=&{3\gamma_1 g_A^4M m_\pi^4 \over (4\pi f_\pi)^4
u^3} \int_0^u {\dif}x\,x^2 \int_{-1}^1{\dif}y \, \Big[ 2uxy +(u^2-x^2y^2)H
\Big] \Big[  (\gamma_1-1)s^2+ \ln(1+s^2) \Big] \nonumber \\ &=&  \gamma_1 M
\Big({g_A m_\pi \over 4\pi f_\pi }\Big)^4 \bigg\{ (\gamma_1-1){8u^4 \over 35}
(11-2\ln 2)+ {8u^2\over 5}(3-\ln2) -{1\over 2}-4u \arctan 2u \nonumber \\ &&
+\bigg({3\over 2}+{1\over 8u^2} \bigg) \ln(1+4u^2) +{3\over 2u^3} \int_0^u
{\dif}x(u^2-x^2) \bigg[(1+u^2-x^2) \ln{1+(u+x)^2\over 1+(u-x)^2} \nonumber \\
 && +4x \Big[\arctan 2x - \arctan(u+x)-\arctan(u-x)\Big]-\ln(1+4x^2) \bigg]
\ln{u+x\over u-x} \bigg\}\,, \end{eqnarray}
with the auxiliary functions $H=\ln(u+x y)- \ln(u-x y)$ and $s=xy +\sqrt{u^2-
x^2+x^2y^2}$. The quantity $s$ has the geometrical meaning of the distance
between a point on a sphere of radius $u$ and an interior point displaced at a
distance $x$ from the center of the sphere. In the same geometrical picture $y$
denotes a directional cosine. Note that iterated three-loop diagrams carry an
energy denominator equal to the difference of nucleon kinetic energies and
therefore they are proportional to the (large) nucleon mass $M=939\,$MeV. 
The additional contributions to $\bar E_n(k_n)$ from the (relativistically
improved) kinetic energy, from $1\pi$-exchange and from iterated $1\pi
$-exchange have been written down in eqs.(32-37) of ref.\cite{nucmat}. In the
case of the $1\pi$-exchange contribution (see eq.(33) in ref.\cite{nucmat}) we
neglect of course the small relativistic $1/M^2$-correction of order ${\cal
O}(k_f^5)$. Note that the parameter $\Gamma_1$ defined in eq.(2) includes also
the cut-off dependent contribution from iterated $1\pi$-exchange (see eq.(39)
in ref.\cite{nucmat}) through its $\gamma_1$-independent piece.   

The full line in Fig.\,2 shows the energy per particle $\bar E_n(k_n)$ of pure
neutron matter as a function of the neutron density $\rho_n =k_n^3/3\pi^2$. 
The dashed line stems from the many-body calculation of the Urbana group
\cite{akmal}. This line should be considered as a representative of the host of
realistic neutron matter calculations \cite{urbana,wiringa} which scatter 
around it. The full line in Fig.\,2 has been obtained by adjusting the 
parameter $\gamma_1$ to the value $\gamma_1 = -0.75$. This number translates 
into a repulsive zero-range $nn$-contact interaction which is comparable in 
strength to the $1\pi$-exchange. (Remember that the contact piece of 
$1\pi$-exchange maps onto $\gamma_1^{(1\pi)}= -1/2$). The other parameter 
$\Gamma_1$ regulating an attractive contribution to $\bar E_n(k_n)$ linear in 
the neutron density $\rho_n$ takes the value $\Gamma_1=2.76$. The cut-off scale
$\Lambda_1 \simeq 0.48\,$GeV which parameterizes (via the relation in eq.(2)) 
this amount of attraction linear in density lies well within the physically 
reasonable range. One observes that the result of realistic neutron matter 
calculation of ref.\cite{akmal} can be reproduced up to quite high neutron 
densities $\rho_n\leq 0.5\,\fmd$. In particular, the downward bending of $\bar 
E_n(k_n)$ above $\rho_n \geq 0.2\fmd$ which showed up in Fig.\,8 of 
ref.\cite{nucmat} (including only pion-exchanges) is now eliminated. Note that 
in the scheme of Lutz et al. \cite{lutz} this downward bending also persisted 
(see Fig.\,6 in ref.\cite{lutzcontra}). Altogether, it seems that most of the 
dynamics underlying the neutron matter equation of state (at moderate densities
$\rho_n \leq 0.5\fmd$) can be covered by pion-exchange and a comparably strong 
short-distance $nn$-repulsion treated up to three-loop order. At low densities,
$\rho_n <0.2\fmd$, the equation of state of pure neutron matter is surprisingly
well approximated by just one half of the kinetic energy, $\bar E_{kin}(k_n)/2=
3k_n^2/20M$, (see the dotted line in Fig.\,2) as emphasized recently in 
ref.\cite{montecarlo}.  

\begin{figure}
\begin{center}
\includegraphics[scale=0.5,clip]{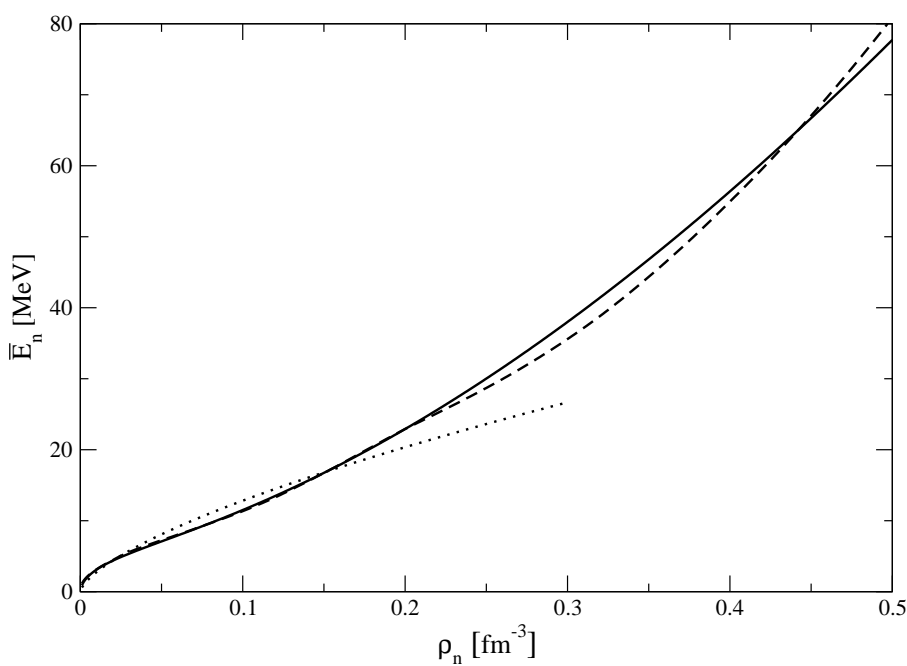}
\end{center}\vspace{-0.2cm}
\caption{The energy per particle $\bar E_n(k_n)$ of pure neutron matter versus
the neutron density $\rho_n = k_n^3/3\pi^2$. The full line is obtained by 
adjusting the strength parameter of the $nn$-contact interaction to $\gamma_1 
=-0.75$ together with a cut-off $\Lambda_1 =0.48\,$GeV. The dashed line stems 
from the many-body calculation of the Urbana group \cite{akmal}. The dotted
line gives one half of the kinetic energy $\bar E_{kin}(k_n)/2 = 3k_n^2/20M$.} 
\end{figure}

\section{Equation of state of isospin-symmetric nuclear matter}
In this section we turn to the equation of state of isospin-symmetric nuclear
matter. In this system with proton-to-neutron ratio equal to one both total 
NN-isospins $I=0,1$ are present and therefore all four parameters 
$\gamma_{0,1}$ and $\Lambda_{0,1}$ come into play. Following the scheme in
sect. 2 we first enumerate the contributions to the energy per particle $\bar 
E(k_f)$ generated by the zero-range NN-contact vertex. The complete term
linear in the nucleon density $\rho= 2k_f^3/3\pi^2$ can again be compactly 
written by introducing a new parameter $\Gamma_0$ which subsumes cut-off
dependent pieces in the $I=0$ channel:
\begin{equation} \bar E(k_f) = - {(\Gamma_0+\Gamma_1) g_A^2 k_f^3 \over 
(4\pi f_\pi)^2} \,, \qquad \Gamma_0 = \gamma_0 +{g_A^2 M \Lambda_0 \over 
(4\pi f_\pi)^2 }(4\gamma_0^2 -4\gamma_0+9) \,.     \end{equation}
The second pion-exchange diagram in Fig.\,1 with two medium insertions on
parallel nucleon lines leads to the following finite (i.e. cut-off 
independent) contribution:   
\begin{equation} \bar E(k_f) = {3(\gamma_0+\gamma_1) g_A^4 M m_\pi^4 \over 10
(8\pi)^3 f_\pi^4} \bigg[22u -{1\over u}-4(5+4u^2) \arctan 2u +\bigg( {1\over 
4u^3} +{5\over u} \bigg) \ln(1+4u^2) \bigg] \,,\end{equation} 
where the meaning of $u$ has changed in this and all following sections to 
$u=k_f/m_\pi$. Diagrams with three medium insertions represent Pauli blocking 
effects in isospin-symmetric nuclear matter. Their contribution to the energy
per particle reads: 
\begin{eqnarray} \bar E(k_f)&=&{9g_A^4 M m_\pi^4 \over 2(4\pi f_\pi)^4
u^3} \int_0^u {\dif}x\,x^2 \int_{-1}^1{\dif}y \,\Big[ 2uxy +(u^2-x^2y^2)H \Big]
\nonumber \\ && \times \Big[(\gamma_0^2+\gamma_1^2-\gamma_0-\gamma_1)s^2 +
(\gamma_0+\gamma_1) \ln(1+s^2)  \Big] \,, \end{eqnarray}
where the linear/quadratic term in $\gamma_{0,1}$ obviously belongs the 
second/third diagram in Fig.\,1. For further reduction of the occurring 
double-integral, see eq.(4). The expansion of the energy per particle up to
order ${\cal O}(k_f^4)$ is completed by adding to the terms eqs.(5,6,7) the
contribution from the (relativistically improved) kinetic energy, (static)
$1\pi$-exchange and iterated $1\pi$-exchange written down in eqs.(5-11) of
ref.\cite{nucmat}.   

Let us first look at generic properties of the nuclear matter equation of state
in our calculation. Binding and saturation occurs in a wide range of the three
relevant combinations of parameters: $\Gamma_0+\Gamma_1$, $\gamma_0+\gamma_1$
and $\gamma^2_0+\gamma^2_1$. We fix the negative minimum of the saturation
curve $\bar E(k_f)$ to the value $-15.3\,$MeV \cite{seeger} by adjusting at
given $(\gamma_0,\gamma_1)$ the strength $\Gamma_0+\Gamma_1$ of the term 
linear in the density $\rho$ appropriately. The horizontal position of each 
minimum determines then the corresponding saturation density $\rho_0(
\gamma_0,\gamma_1)=\rho_0(\gamma_1,\gamma_0)$. It is obviously symmetric under 
the exchange of the two parameters $\gamma_0 \leftrightarrow \gamma_1$. In 
Fig.\,3, we show contours of constant saturation density $\rho_0$ in the 
$(\gamma_0,\gamma_1)$-plane. To a good approximation these contours are 
concentric circles around a point in parameter space where the saturation 
density becomes maximal. We find numerically the following approximate 
parameter dependence of the saturation density (in units of $\fmd$):
\begin{equation} \rho_0[\fmd\,] = 0.191-0.0388((\gamma_0-0.387)^2+(\gamma_1-
0.387)^2)\,. \end{equation}
Note that the maximal saturation density $\rho_0^{\rm max} = 0.191\fmd$ 
possible in the present perturbative calculation lies just at the upper end of
the range $\rho_0= (0.166\pm 0.027)\fmd$ quoted in 
ref.\cite{blaizot}. Interestingly, the maximal saturation density $\rho_0^{
\rm max} = 0.191\fmd$ is reached in the near vicinity of the special point 
$\gamma_0=\gamma_1 =1/2$. The latter describes the situation in which the
(explicitly introduced) short-range NN-interaction exactly cancels the contact
pieces of the $1\pi$-exchange. The additional constraint $\gamma_1 = -0.75$
taken over from the fit to the neutron matter equation of state (see sect.\,2) 
leads (together with $\gamma_0 = 0.39$) however to a somewhat too low 
saturation density of $\rho_0= 0.134\fmd$. The other parameter $\Gamma_0=5.63$
needed to keep the depth of the saturation curve at $-15.3\,$MeV translates 
into an isospin-0 cut-off of $\Lambda_0 =0.55\,$GeV. Its value does not differ 
much from the isospin-1 cut-off $\Lambda_1 \simeq 0.48\,$GeV needed for the
neutron  matter equation of state and it also lies within the physically 
acceptable range. Note that momentum cut-offs around $0.5\,$GeV have also been
used in chiral perturbation theory calculations of NN-phase shifts in 
refs.\cite{epel,mach}. In these works the cut-off serves a somewhat different 
purpose of regulating the (non-perturbative) Lippmann-Schwinger equation. 
However, no comparable fine-tuning of $\Lambda$ is needed in these works since
the cut-off dependence of NN-observables can to a large extent be balanced by 
the strengths of NN-contact terms. 

\begin{figure}
\begin{center}
\includegraphics[scale=0.85,clip]{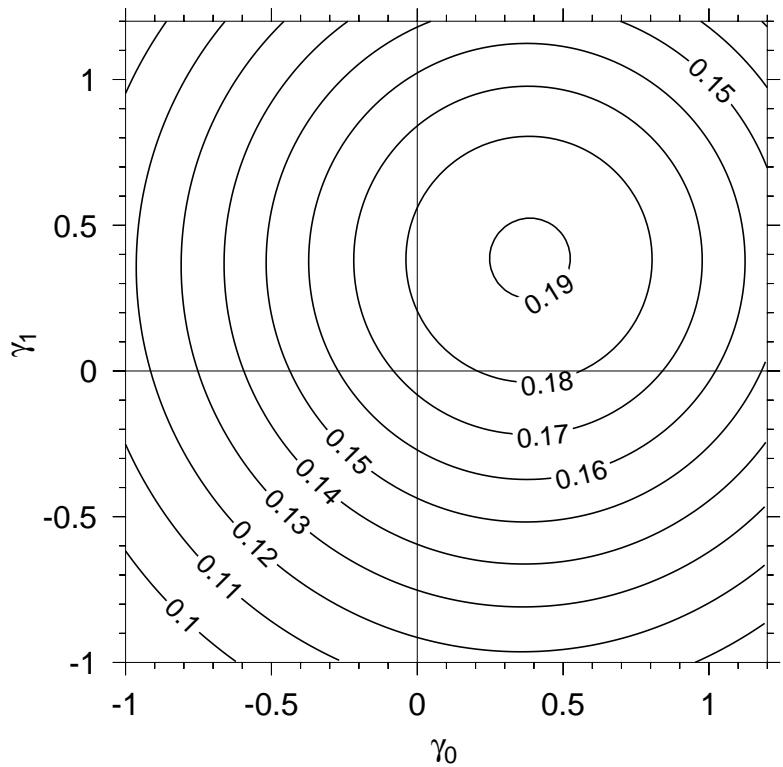}
\end{center}\vspace{-0.5cm}
\caption{Contours of constant nuclear matter saturation density $\rho_0$ (in
units of $\fmd$) in the $(\gamma_0,\gamma_1)$-plane of short-range parameters. 
The maximal saturation density is $\rho_0^{\rm max}=0.191\,\fmd\simeq 1.15
\rho_0^{\rm emp}$. The binding energy per particle is fixed to the value $-\bar
E(k_{f0}) = 15.3\,$MeV.}
\end{figure}
\begin{figure}
\begin{center}
\includegraphics[scale=0.85,clip]{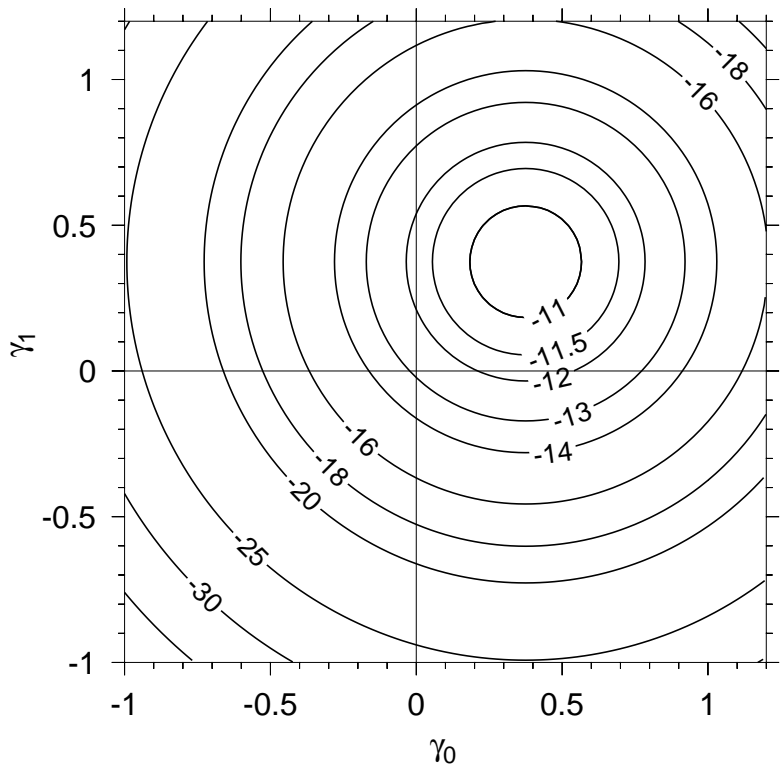}
\end{center}\vspace{-0.5cm}
\caption{Contours of constant saturation energy $\bar E(k_{f0})$ (in units of
MeV) in the $(\gamma_0,\gamma_1)$-plane of short-range parameters. The maximal 
saturation energy is $\bar E(k_{f0})^{\rm max}=-10.7\,$MeV. The nuclear matter
saturation density is fixed to the value $\rho_0=0.158\fmd$ \cite{sick}.}
\end{figure}

One may also invert the fitting procedure and extract this way the parameter 
dependence of the maximal binding energy $-\bar E(k_{f0})$ per nucleon at
fixed saturation density $\rho_0$. For this purpose we choose the value
$\rho_0=0.158\fmd$ as obtained by extrapolation from inelastic electron
scattering off heavy nuclei \cite{sick}. Fig.\,4 shows the resulting contours
of constant saturation energy $\bar E(k_{f0})$ in the $(\gamma_0, \gamma_1)
$-plane of short-range parameters. Again, to a good approximation these
contours are concentric circles around the point $\gamma_0=\gamma_1=0.38$ in
parameter space where the saturation energy becomes maximal, $\bar
E(k_{f0})^{\rm max}= -10.7\,$MeV.

The full curve in Fig.\,5 shows the equation of state of isospin-symmetric
nuclear matter corresponding to the parameter set $(\gamma_0 = 0.39,\, \gamma_1
= -0.75,\, \Gamma_0=5.63,\,\Gamma_1=2.76)$ constrained by the equation of state
of pure neutron matter. The nuclear matter compressibility (related to the 
curvature of the saturation curve at its minimum) comes out as $K=k_{f0}^2 \bar
E''(k_{f0})=243\,$MeV, in good agreement with the presently accepted empirical 
value $K = (250\pm 25)\,$MeV \cite{blaizot,vretenar}. For comparison, the value
$K = 253\,$MeV \cite{lutzcontra} of the nuclear matter compressibility is 
obtained if the (explicit) short-range NN-terms are switched off $(\gamma_0 =
\gamma_1= 0,\,\Lambda_0=\Lambda_1 =0.61\,$GeV). The dashed line in Fig.\,5
shows the equation of state $\bar E(k_f)$ with maximal saturation density
$\rho_0^{\rm max}=0.191\fmd$ which is realized by the parameter choice: 
$\gamma_0 = \gamma_1= 0.39$ and $\Gamma_0 + \Gamma_1= 6.46$ (or equivalently
$\Lambda_0=\Lambda_1 = 0.60\,$GeV). In this case the nuclear matter 
compressibility changes only slightly to $K=250\,$MeV. The dotted line in 
Fig.\,5 stems from the sophisticated many-body calculation of the Urbana group 
\cite{akmal} using the Argonne V18 NN-potential plus boost corrections and 
three-nucleon interactions. By giving up the constraints from the neutron 
matter equation of state one could also reproduce this curve (for moderate 
densities $\rho\leq 0.3 \fmd$). The corresponding best fit parameters would be 
$\gamma_0=\gamma_1=-0.19$ and $\Lambda_0=\Lambda_1=0.59\, $GeV. Altogether, it 
seems that most of the parameter freedom of our calculation is already 
exhausted by adjusting the nuclear matter saturation point. The compressibility
is rather stable and takes on reasonable values around $K\simeq 250\,$MeV.   

\bigskip

\begin{figure}
\begin{center}
\includegraphics[scale=0.5,clip]{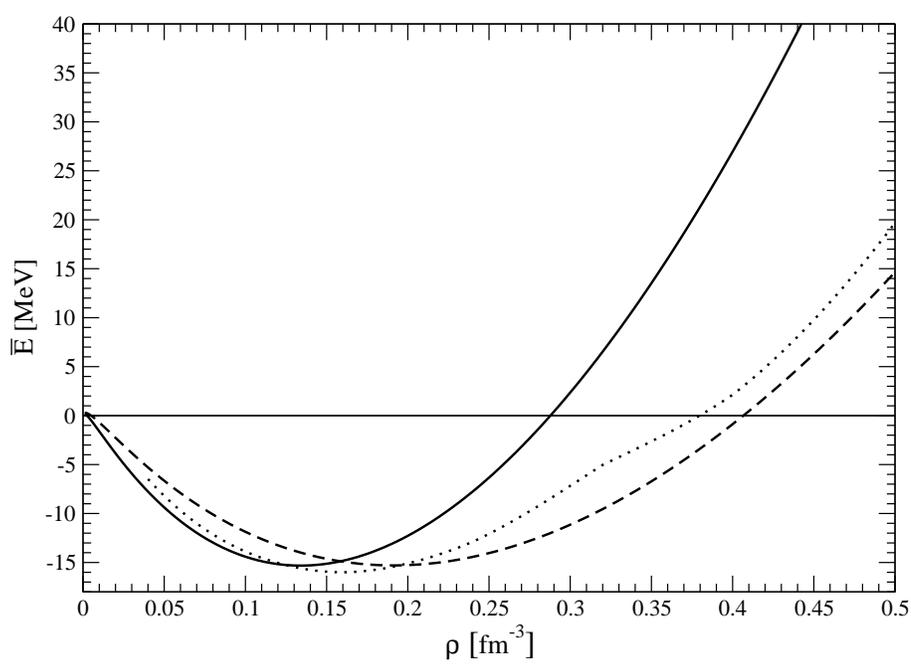}
\end{center}\vspace{-0.2cm}
\caption{The energy per particle $\bar E(k_f)$ of isospin-symmetric nuclear 
matter versus the nucleon density $\rho=2k_f^3/3\pi^2$. The full curve
corresponds to the parameter set constrained by the neutron matter equation of
state. The dashed curve shows the case of maximal saturation density
$\rho_0^{\rm max}=0.191\,\fmd$. The dotted line stems from the many-body
calculation of the Urbana group \cite{akmal}.}  
\end{figure}

\section{Real single-particle potential}
Next, we turn to the real part of the single-particle potential $U(p,k_f)$ 
below the Fermi surface ($p<k_f$). As outlined in ref.\cite{pot}, the
contributions to the (real) nuclear mean-field $U(p,k_f)$ can be classified as
two-body and three-body potentials. The first diagram in Fig.\,1 (combined with
all cut-off dependent terms from iterated diagrams) leads to a momentum
independent contribution to the two-body potential of the form:   
\begin{equation} U_2(p,k_f) = - {(\Gamma_0+\Gamma_1) g_A^2 k_f^3 \over 2(2\pi 
f_\pi)^2} \,, \end{equation}
which is just twice the energy per particle written in eq.(5). From the second
pion-exchange diagram in Fig.\,1 one derives a (finite) contribution to the
two-body potential of the form: 
\begin{eqnarray} U_2(p,k_f) &=& {(\gamma_0+\gamma_1) g_A^4 M m_\pi^4 \over 
(4\pi)^3  f_\pi^4} \bigg\{u+{1\over 8x}(1+3u^2-3x^2) \ln{1+(u+x)^2 \over 1+
(u-x)^2} \nonumber \\ && +{1\over 4x}(x^3-3x-3u^2x-2u^3)\arctan(u+x) \nonumber 
\\ && +{1\over 4x} (x^3-3x-3u^2x+2u^3) \arctan(u-x) \bigg\}\,, \end{eqnarray}
with the abbreviation $x=p/m_\pi$. The second and third diagram in Fig.\,1 give
each rise to three different contributions to the three-body potential. 
Altogether, they read:
\begin{eqnarray} U_3(p,k_f) &=& {3(\gamma_0+\gamma_1) g_A^4 M m_\pi^4 \over 
(4\pi f_\pi)^4} \int_{-1}^1{\dif}y \, \bigg\{ \Big[uxy +(u^2-x^2y^2){H\over 2}
\Big] \Big[(r-1)s^2+ \ln(1+s^2)\Big]\nonumber \\ & &  +\int_{-xy}^{s-xy}{\dif} 
\xi\,\bigg[2u \xi+(u^2-\xi^2) \ln{u+\xi \over u-\xi}\bigg] \,{ r(xy+\xi)+(r-1) 
(xy+\xi)^3 \over 1+(xy+\xi)^2} \nonumber \\ & & +\int_0^u {\dif} \xi\, {\xi^2 
\over x} \Big[ (r-1)\sigma^2 +\ln(1+\sigma^2) \Big]\ln{|x+\xi y|\over|x-\xi y|}
\bigg\}\,,  \end{eqnarray}
with the auxiliary function $\sigma=\xi y +\sqrt{u^2-\xi^2+\xi^2y^2}$ and the
ratio of parameters $r=(\gamma_0^2+\gamma_1^2)/(\gamma_0+\gamma_1)$. The real
single-particle potential $U(p,k_f)$ is completed by adding to the terms
eqs.(9,10,11) the contributions from $1\pi$-exchange and iterated
$1\pi$-exchange written down in eqs.(8-13) of ref.\cite{pot}. Again, the higher
order relativistic $1/M^2$-correction to $1\pi$-exchange (see eq.(8) in
ref.\cite{pot}) is neglected here for reasons of consistency.

The slope of the real single-particle potential $U(p,k_f)$ at the Fermi 
surface $p=k_f$ determines the effective nucleon mass (the product of
"$k$-mass" and "$E$-mass" divided by the free mass in the nomenclature of 
ref.\cite{mahaux}) via a relation $M^*(k_f)= M/(1+B-k_f^2/2M^2)$. For the 
slope parameter $B$ we find the following numerical result (choosing
$k_f=2m_\pi$):
\begin{equation} B= {M \over p}{\partial U(p,k_f)\over \partial p}\bigg|_{p=
k_f= 2m_\pi} = -0.239-1.245 ((\gamma_0-0.388)^2+(\gamma_1-0.388)^2)
\,.\end{equation} 
The negative coefficient in front of the quadratic polynomial in $\gamma_0$ 
and $\gamma_1$ is known analytically from Galitskii's calculation 
\cite{galitski,fetter} and it reads in our notation: $g_A^4 M^2 k_f^2 (1-7\ln2)
/(80\pi^4 f_\pi^4)$. One deduces from eq.(12) that the slope parameter $B$ is
negative definite. This implies that according to our calculation the effective
nucleon mass at the Fermi-surface $M^*(k_{f0})$ is always larger than the free
nucleon mass $M=939\,$MeV. While the iteration of $1\pi$-exchange with the
NN-contact vertex (second diagram in Fig.\,1 being linear in $\gamma_0+
\gamma_1$) can of course reduce the effective nucleon mass, there are
counteracting quadratic effects from the iteration of the NN-contact
interaction with itself, which set a limit to this. The lower bound on the
effective nucleon mass in the present calculation is $M^*(k_{f0}) \geq
1.4M$. It exceeds considerably the value $M^*(k_{f0}) \simeq 1.15M$ found in
the self-consistent Brueckner calculation of ref.\cite{grange} (see Fig.\,6
therein). Interestingly, the parameter set ($\gamma_0=\gamma_1=0.39$) leading
to the minimal effective nucleon mass coincides with that of maximal saturation
density $\rho_0^{\rm max}=0.191\fmd$. Both features occur simultaneously in a
situation where the (explicitly introduced) short-range NN-interaction cancels
almost completely the contact piece of the $1\pi$-exchange.    

\bigskip

\begin{figure}
\begin{center}
\includegraphics[scale=0.5,clip]{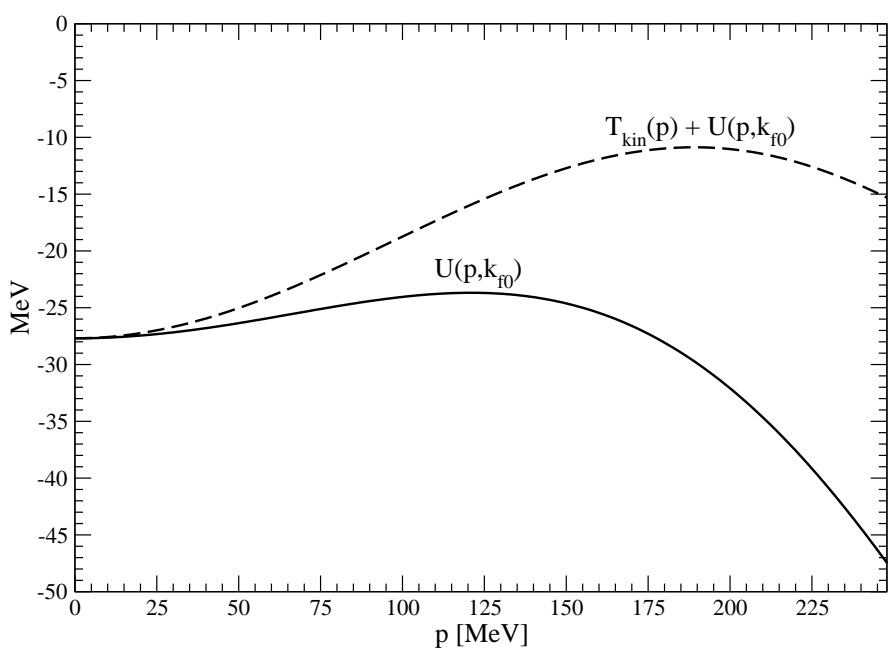}
\end{center}\vspace{-0.2cm}
\caption{The lower curve shows the real part of the single-particle potential
$U(p,k_{f0})$ at saturation density $k_{f0}= 248\,$MeV for the parameter set
constrained by the equation of state of pure neutron matter. The upper curve
includes the relativistically improved kinetic energy $T_{\rm kin}(p) = p^2/2M-
p^4/8M^3$.}
\end{figure}

The lower curve in Fig.\,6 shows the real part of the single-particle potential
$U(p,k_{f0})$ at saturation density $k_{f0}= 248\,$MeV for the parameter set
constrained by the equation of state of pure neutron matter. The potential 
depth of $U(0,k_{f0})=-27.7$\,MeV is by far too weakly attractive, i.e. at
least a factor of two too small. Moreover, a very strong decrease of $U(0,
k_{f0})$ sets in above $p=k_{f0}/2= 124\,$MeV. The consequence of this behavior
is that the total single-particle energy $T_{\rm kin}(p)+U(p,k_{f0})$, shown by
the dashed line in Fig.\,6, does not rise monotonically with the nucleon
momentum $p$. From that perspective the single-particle potential $U(p,k_{f0})$
presented in Fig.\,6 is as unrealistic as the one in the scheme of Lutz et al. 
\cite{lutz,lutzcontra}.  The requirements of a good neutron matter equation of
state and good single-particle properties in isospin-symmetric nuclear matter
cannot be fulfilled simultaneously in a calculation to fourth order. A possible
remedy of this could be a $p^2$-dependent NN-contact interaction (contributing
at fifth order in small momenta).  

\bigskip

\begin{figure}
\begin{center}
\includegraphics[scale=0.5,clip]{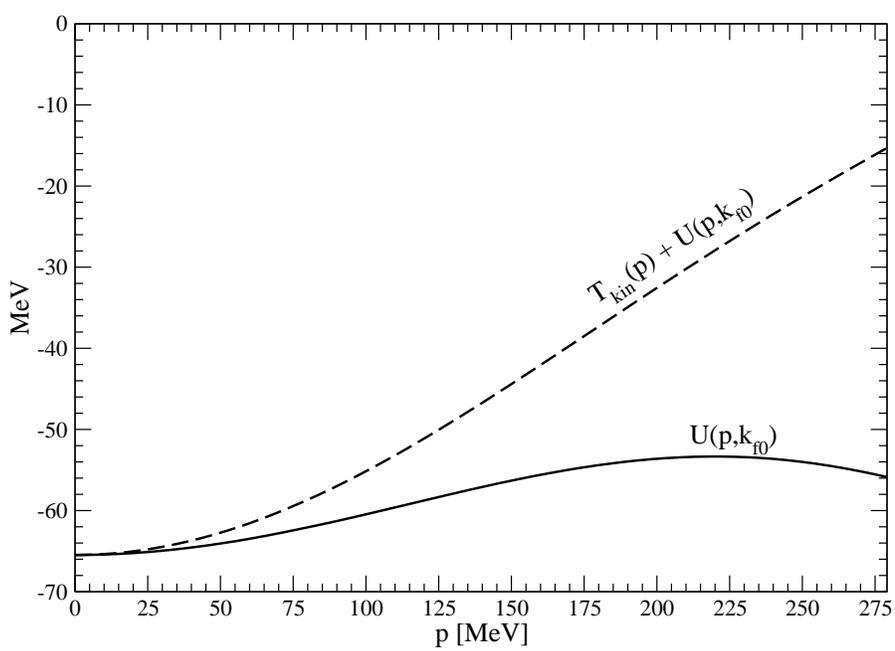}
\end{center}\vspace{-0.2cm}
\caption{The lower curve shows the real part of the single-particle potential
$U(p,k_{f0})$ at $k_{f0}= 279\,$MeV for the case of maximal saturation density 
$\rho_0^{\rm max}= 0.191\fmd$. The upper curve includes the relativistically 
improved kinetic energy $T_{\rm kin}(p) = p^2/2M- p^4/8M^3$.}
\end{figure}

The lower curve in Fig.\,7 shows the real part of the single-particle potential
$U(p,k_{f0})$ for the case of maximal saturation density $\rho_0^{\rm max}= 
0.191\fmd$. The potential depth of $U(p,k_{f0})=-65.5\,$MeV is now close to 
the value $-64\,$MeV obtained in the calculation of ref.\cite{grange} using the
Paris NN-potential. The momentum dependence is also reasonable since its leads
to a monotonic rise of the total single-particle energy $T_{\rm kin}(p)+
U(p,k_{f0})$, as it should be. Note that both dashed curves in Figs.\,6,7 end
at the Fermi surface $p=k_{f0}$ with the value $\bar E(k_{f0})=-15.3\,$MeV as
required by the Hugenholtz-Van-Hove theorem \cite{vanhove}. This serves an
important check on our (analytical and numerical) calculations.  
    
\section{Imaginary single-particle potential}

Next, we come to the imaginary single-particle potential $W(p,k_f)$ as it
arises  from the iteration of the NN-contact interaction with $1\pi$-exchange
and with itself. For $p\leq k_f$ the imaginary single-particle potential
$W(p,k_f)$ is equal to the half-width of nucleon hole states in the Fermi
sea. As outlined in ref.\cite{pot} the contributions to $W(p,k_f)$ can be
classified as two-body, three-body and four-body terms. From a technical point
of view a $\nu$-body term involves an integral over the product of $\nu-1$
Fermi spheres of radius $k_f$. The second and third diagram in Fig.\,1 give
rise to a two-body term of the form:    
\begin{eqnarray} W_2(p,k_f) &=&{(\gamma_0+\gamma_1)g_A^4Mm_\pi^4\over(8\pi)^3 
f_\pi^4} \bigg\{(r-1) \bigg( {x^4 \over 5}-2u^2x^2 -3u^4\bigg)+ 4+14u^2-{22x^2 
\over 3}\nonumber \\ & &  +{2\over x} (3x^2-3u^2-1) \Big[\arctan(u+x)-
\arctan(u-x) \Big] \nonumber \\ & &  +{1\over x} (x^3-3u^2x-3x-2u^3)
\ln[1+(u+x)^2]\nonumber \\ & &  +{1\over x} (x^3-3u^2x-3x+2u^3) \ln[1+(u-x)^2]
\bigg\}\,, \end{eqnarray}
with the coefficient $r$ defined after eq.(11). The associated three-body term
reads:
\begin{eqnarray} W_3(p,k_f) &=& {(\gamma_0+\gamma_1)g_A^4Mm_\pi^4\over(8\pi)^3
f_\pi^4}\bigg\{4(r-1)(3u^4-x^4) +{41x^2\over 3}-31u^2-5-(u^2-x^2)^2\nonumber 
\\ & & -3\ln(1+4x^2)+\bigg(6x-{3\over 2x}\bigg)\arctan2x   
+\Big[\arctan(u+x)- \arctan(u-x) \Big] \nonumber \\ & & \times {1\over 2x}\Big[
(u^2-x^2)^3 +(12u^2+27) (u^2-x^2)+8\Big]  \nonumber \\ & & +\Big( 6+9u^2-3x^2+
{2u^3\over x}\Big)\ln[1+(u+x)^2]  \nonumber \\ & & + \Big( 6+9u^2-3x^2-{2u^3
\over x} \Big) \ln[1+(u-x)^2] \bigg\}\,, \end{eqnarray} 
and the four-body term is given by the expression:
\begin{eqnarray} W_4(p,k_f) &=&{(\gamma_0+\gamma_1)g_A^4Mm_\pi^4\over (8\pi)^3
f_\pi^4}\bigg\{2(r-1) \bigg({17 x^4 \over 5}-3u^4-2u^2x^2\bigg)+1+20u^2-{28x^2 
\over 3}\nonumber \\ & & +6\ln(1+4x^2)+\bigg({3\over x}-12x\bigg)
\arctan2x +\Big[\arctan(u+x)- \arctan(u-x) \Big] \nonumber \\ & & \times 
{1\over 2x}\Big[(x^2-u^2)^3+3x^4+6u^2x^2-9u^4+27x^2-15u^2-7\Big]  \nonumber \\ 
& &+2(x^2-3u^2-3)\ln\Big([1+(u+x)^2][1+(u-x)^2]\Big) +(u^2-x^2)^2\bigg\}\,.
\end{eqnarray}
The additional contributions to $W(p,k_f)$ from the iterated $1\pi$-exchange
Hartree and Fock diagrams are collected in eqs.(20-25) of ref.\cite{pot}. In
the appendix we present also novel analytical expressions for the three-body 
and four-body Hartree term of iterated $1\pi$-exchange. The total imaginary 
single-particle potential evaluated at zero nucleon momentum $(p=0)$ can even 
be written as a closed form expression:     
\begin{eqnarray}  W(0,k_f) &=&{3\pi g_A^4M m_\pi^4 \over (4\pi f_\pi)^4}\bigg\{
\Big[3+2(\gamma_0^2+\gamma_1^2-\gamma_0-\gamma_1)\Big]{u^4\over 4}+(\gamma_0+
\gamma_1-3) u^2-{2u^2\over 1+u^2}+{\pi^2\over 12}\nonumber\\ && +{\rm Li}_2(
-1-u^2) +\bigg[5-\gamma_0-\gamma_1+\ln(2+u^2)-{1\over 2}
\ln(1+u^2) \bigg] \ln(1+u^2) \bigg\} \,, \end{eqnarray}
where Li$_2(-a^{-1}) = \int_0^1 d\zeta \, (\zeta+a)^{-1} \ln\zeta$ denotes the 
conventional dilogarithmic function. Note that eq.(16) determines the density
dependence of the half-width of nucleon hole states at the bottom of the Fermi 
sea. 

\begin{figure}
\begin{center}
\includegraphics[scale=0.5,clip]{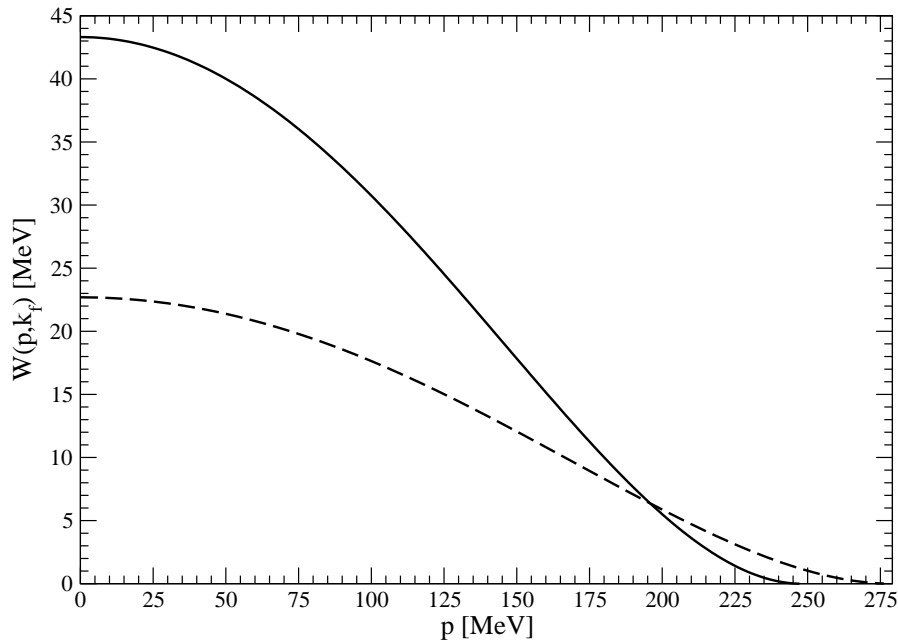}
\end{center}\vspace{-0.2cm}
\caption{The imaginary single-particle potential $W(p,k_{f0})$ at saturation 
density versus the nucleon momentum $p$. The full curve corresponds to the 
parameter set constrained by the neutron matter equation of state. The dashed 
curve shows the case of maximal saturation density $\rho_0^{\rm max}=0.191\,
\fmd$.}  
\end{figure}

Fig.\,8 shows the imaginary single-particle potential $W(p,k_{f0})$ at 
saturation density as a function of the nucleon momentum $p$. The full curve 
corresponds to the parameter set constrained by the neutron matter equation of
state (where $k_{f0}=248\,$MeV). The resulting value $W(0,k_{f0})=43.3\,$MeV
is somewhat larger than the value $W(0,k_{f0})\simeq 40\,$MeV obtained in the
self-consistent Brueckner calculation of ref.\cite{grange} using the
phenomenological Paris NN-potential. The dashed curve in Fig.\,8 shows the 
case of maximal saturation density $\rho_0^{\rm max}=0.191\,\fmd$. With 
associated parameters $\gamma_0=\gamma_1=0.39$ the imaginary single-particle 
potential at $p=0$ comes out as $W(0,k_{f0})=22.7\,$MeV. This value is close to
the result $W(0,k_{f0})\simeq 20\,$MeV found in the calculation of 
ref.\cite{schuck} employing the Gogny D1 effective interaction. For comparison
the value $W(0,k_{f0})=29.7\,$MeV \cite{pot} results from the pure 
pion-exchange dynamics. The momentum dependence of $W(p,k_{f0})$ is generic. As
a consequence of the decreasing phase space available for redistributing the
nucleon-hole state energy the curves in Fig.\,8 drop with momentum $p$ and
$W(p,k_{f0})$ reaches zero at the Fermi surface $p=k_{f0}$. According to
Luttinger's theorem \cite{luttinger} this vanishing is of quadratic order,
$W(p,k_f)\sim(p-k_f)^2$, a feature which is clearly exhibited by the curves in
Fig.\,8.

\section{Asymmetry energy}
This section is devoted to the density dependent asymmetry energy $A(k_f)$. The
asymmetry energy is generally defined by the expansion of the energy per
particle of isospin-asymmetric nuclear matter (described by different proton
and neutron Fermi momenta $k_{p,n} = k_f(1\mp \delta)^{1/3}$) around the
symmetry line: $\bar E_{as}(k_p,k_n) = \bar E(k_f) + \delta^2\, A(k_f) + {\cal
O}(\delta^4)$. Evaluating the first diagram in Fig.\,1 and combining that
partial result with all cut-off dependent terms from iterated diagrams one
obtains the following contribution (linear in density) to the asymmetry energy:
\begin{equation} A(k_f) = {g_A^2 k_f^3 \over 3(4\pi f_\pi)^2}(3\Gamma_0-
\Gamma_1)   \,, \end{equation}
with the parameters $\Gamma_0$ and $\Gamma_1$ defined in eqs.(2,5). 
From the second pion-exchange diagram in Fig.\,1 with two medium insertions on 
parallel nucleon lines one gets a (finite) contribution to the
asymmetry energy of the form: 
\begin{eqnarray} A(k_f) &=& {g_A^4 M m_\pi^4 \over 3(8\pi)^3 f_\pi^4} \bigg\{ 
2(\gamma_0+\gamma_1)u +16\gamma_0u^2 \arctan 2u \nonumber \\ && -\bigg[(6
\gamma_0+2\gamma_1) u + {\gamma_0+\gamma_1 \over 2u}\bigg] \ln(1+4u^2) \bigg\}
\,.\end{eqnarray}
The same (pion-exchange) diagram with three medium insertions gives rise to a
contribution to the asymmetry energy which can be represented as a
double-integral:  
\begin{eqnarray} A(k_f) &=& {g_A^4 M m_\pi^4 \over (4\pi f_\pi)^4 u^3}
\int_0^u {\dif} x\,x^2\int_{-1}^1{\dif} y\,\Bigg\{(\gamma_0+\gamma_1)\bigg\{
\bigg[{uxy (11u^2-15x^2y^2)\over 3(u^2-x^2y^2)} +(u^2-5x^2y^2){H\over 2} \bigg]
\nonumber\\ && \times \Big[\ln(1+s^2)-s^2\Big]+{4u^2s^4H \over 3(1+s^2)}+
{s^2\over 6(1+s^2)^2}\Big[2uxy +(u^2-x^2y^2)H \Big] \nonumber\\ && \times\Big[
8s(1+s^2) (5s'-3s-s'') +(3+s^2) (8ss'-3s^2-8s'\,^2) \Big]\bigg\} \nonumber\\ &&
+4\gamma_1u^2\bigg[{2uxy[\ln(1+s^2)-s^2]
\over 3(u^2-x^2y^2)} +\bigg(\ln(1+s^2)-{s^2(3+5s^2)\over 3(1+s^2)}\bigg)H\bigg]
\Bigg\}  \,, \end{eqnarray}
with $s' = u\, \partial s/\partial u$ and $s''=u^2\,\partial^2 s/\partial u^2$
denoting partial derivatives. In the case of the third diagram in Fig.\,1 with 
three medium insertions the occurring double-integral can actually be solved.
The $k_f^4$-contribution to the asymmetry energy originating from the iterated 
NN-contact interaction reads:  
\begin{equation} A(k_f) = \Big({g_A k_f \over 2\pi f_\pi}\Big)^4 {M \over 15}
\Big[ \gamma_1^2 (3-\ln2 ) -\gamma_0^2 (2+\ln2 ) \Big] \,. \end{equation}
The absence of an interference term proportional to $\gamma_0 \gamma_1$ is 
highly remarkable. The asymmetry energy $A(k_f)$ is completed by adding to the
terms in eqs.(17-20) the contributions from the (relativistically improved) 
kinetic energy, (static) $1\pi$-exchange and iterated $1\pi$-exchange written
down in eqs.(20-26) of ref.\cite{nucmat}.   

\begin{figure}
\begin{center}
\includegraphics[scale=0.5,clip]{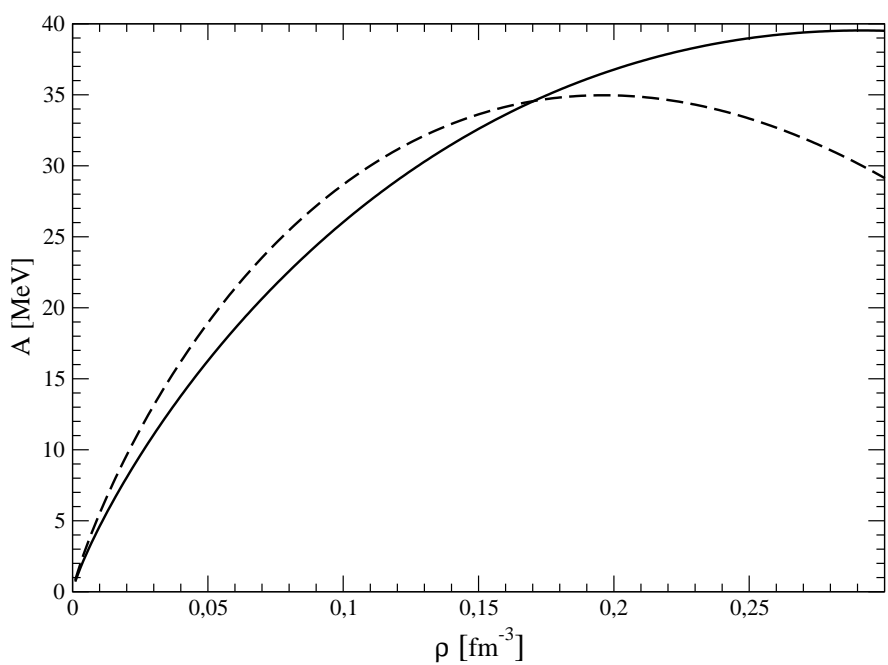}
\end{center}\vspace{-0.2cm}
\caption{The asymmetry energy $A(k_f)$ versus the nucleon density $\rho=2k_f^3
/3\pi^2$. The full curve corresponds to the parameter set constrained by the 
neutron matter equation of state. The dashed curve shows the case of maximal 
saturation density $\rho_0^{\rm max}=0.191\,\fmd$.}  
\end{figure}

In Fig.\,9, we show the asymmetry energy $A(k_f)$ versus the nucleon density 
$\rho=2k_f^3 /3\pi^2$. The full curve corresponds to the parameter set 
constrained by the neutron matter equation of state. The predicted value of the
asymmetry energy at saturation density $\rho_0=0.134\fmd$ is $A(k_{f0})=
30.8\,$MeV. This is close to the empirical value $A(k_{f0})\simeq 33\,$MeV of
ref.\cite{seeger} obtained in extensive fits to nuclide masses.  As a 
consequence of the built in constraints from the neutron matter equation of
state the full curve in Fig.\,9 starts to bend downwards only above relatively 
high densities $\rho > 0.3\fmd$. The dashed curve in Fig.\,9 shows the case of 
maximal saturation density $\rho_0^{\rm max}= 0.191\,\fmd$ realized by the
parameters $\gamma_0=\gamma_1 = 0.39$. We use furthermore the 
decomposition\footnote{Returning to eq.(2) one notices that $\Gamma_1$ and
$\gamma_1$ are necessarily approximately equal if $\gamma_1 \approx 1/2$.}
$\Gamma_0+ \Gamma_1= 6.04+ 0.42$ as it follows from choosing a common cut-off
$\Lambda_0=\Lambda_1 = 0.60\,$GeV. The resulting value $A(k_{f0})=35.0\,$MeV of
the asymmetry energy at saturation density $\rho_0^{\rm max}=0.191\fmd$ is also
in good agreement with the empirical value. However, in that case the downward 
bending of $A(k_f)$ sets in immediately above saturation density. The situation
concerning the high density behavior of the asymmetry energy is still
controversial. While Brueckner-Hartree-Fock calculations of ref.\cite{bombaci}
predict a monotonic increase with density the results of the Urbana group
\cite{wiringa} show a downward bending of $A(k_f)$ at densities $\rho =0.4\fmd$
or higher (see Fig.\,8 therein). Altogether, it seems that the value of the 
asymmetry at saturation density is rather stable while the detailed density 
dependence of $A(k_f)$ reflects more pronouncedly some properties of the 
corresponding neutron matter equation of state.

\section{Nuclear matter at finite temperatures}
In this section, we discuss nuclear matter at finite temperatures $T\leq 
30\,$MeV. We are particularly interested in the first-order liquid-gas phase 
transition of isospin-symmetric nuclear matter and its associated critical 
point $(\rho_c,T_c)$. As outlined in ref.\cite{liquidgas} a thermodynamically 
consistent extension of the present (perturbative) calculational scheme to 
finite temperatures is to relate it directly to the free energy per particle 
$\bar F(\rho,T)$, whose natural thermodynamical variables are the nucleon 
density $\rho$ and the temperature $T$. In that case the free energy density 
$\rho\bar F(\rho,T)$ of isospin-symmetric nuclear matter consists of a sum of 
convolution integrals over interaction kernels ${\cal K}_j$ multiplied by 
powers of the density of nucleon states in momentum space:
\begin{equation} 
d(p_j) = {p_j\over 2\pi^2} \bigg[ 1+\exp{p_j^2 -2M \tilde \mu
\over 2M T} \bigg]^{-1} \,. \end{equation}
The effective one-body "chemical potential" $\tilde \mu(\rho,T)$ entering the
Fermi-Dirac distribution in eq.(21) is determined by the relation to the 
particle density $\rho = 4\int_0^\infty {\dif}p_1\,p_1 d(p_1)$. We now 
write down the additional interaction kernels arising from the NN-contact 
interaction proportional to $\gamma_{0,1}$. The contact interaction at first 
order and the cut-off dependent pieces from iterated diagrams are summed up via
the parameters $\Gamma_{0,1}$. Their contribution to the two-body kernel reads:
\begin{equation}{\cal K}_2^{(\Gamma)} =-{3g_A^2 \over 2 f_\pi^2 }(\Gamma_0+
\Gamma_1) p_1p_2\,.  \end{equation}
Obviously, this kernel generates a temperature independent contribution to the 
free energy per particle which just grows linear in density, $\bar F(\rho,T) 
\sim \rho$. From the second pion-exchange diagram in Fig.\,1 one derives a
further (finite) contribution to the two-body kernel of the form: 
\begin{eqnarray} 
{\cal K}_2^{(\gamma)} &=& {3(\gamma_0+\gamma_1)g_A^4M m_\pi^2 \over 64\pi 
f_\pi^4}  \bigg\{ m_\pi \ln{m_\pi^2+(p_1+p_2)^2 \over m_\pi^2+(p_1-p_2)^2}
\nonumber \\ && -2(p_1+p_2) \arctan{p_1+p_2\over m_\pi} + 2(p_1-p_2) 
\arctan{p_1-p_2\over m_\pi}  \bigg\} \,. \end{eqnarray} 
Temperature and density Pauli blocking effects in intermediate NN-states are 
incorporated in the three-body kernel ${\cal K}_3$. The second and third 
diagram in Fig.\,1 together give rise to the following contribution to the
three-body kernel:   
\begin{equation} {\cal K}_3^{(\gamma)} ={3 g_A^4M  \over8 f_\pi^4 } \int_{|p_1-
p_2|}^{p_1+p_2} \!\!{\dif} q\,\bigg[\gamma_0^2+\gamma_1^2 -{(\gamma_0+\gamma_1)
q^2\over m_\pi^2 +q^2} \bigg]\ln {|p_1^2-p_2^2+q^2+
2p_3 q| \over |p_1^2-p_2^2+q^2-2p_3 q|} \,.  \end{equation}
The remaining kernels building up the free nucleon gas part and the interaction
contributions from $1\pi$-exchange and iterated $1\pi$-exchange are written
down in eqs.(4,5,6,7,11,12) of ref.\cite{liquidgas}. Again, the (higher order)
relativistic $1/M^2$-correction to the $1\pi$-exchange two-body kernel ${\cal 
K}_2^{(1\pi)}$ (see eq.(5) in ref.\cite{liquidgas}) is neglected here for
reasons of consistency. A special feature of the finite temperature formalism 
is the so-called anomalous contribution $\rho \bar {\cal A}(\rho,T)$ with no 
counterpart in the calculation to the ground state energy density $\rho \bar
E(k_f)$ at $T=0$ \cite{thouless,kohn}. The basic physical mechanism behind this
"anomalous" contribution is the possibility of interactions between
particles and holes of the same momentum at finite temperature. From the
combination of pion-exchange and zero-range NN-contact interaction we derive
the following anomalous contribution to the free energy per particle of
isospin-symmetric nuclear matter:    
\begin{eqnarray} \bar{\cal A}(\rho,T) &=& - { [\Omega_1'(\rho,T)]^2 \over
2\rho\, \Omega_0'' (\rho,T)}\nonumber \\ && + {9 g_A^4 \over 8 f_\pi^4 T \rho} 
\int_0^\infty \!\!{\dif} p_1\int_0^\infty \!\!{\dif} p_2\int_0^\infty \!\!
{\dif} p_3\,\, d(p_1)d(p_2)[2\pi^2 d(p_2)-p_2] d(p_3)\nonumber \\ && \times
\bigg[(\gamma_0+\gamma_1-1)p_1+ {m_\pi^2\over 4p_2} \ln{m_\pi^2+(p_1+p_2)^2
\over m_\pi^2+(p_1-p_2)^2}\bigg] \nonumber \\ && \times  \bigg[(\gamma_0+
\gamma_1-1)  p_3+{m_\pi^2\over 4p_2}\ln{m_\pi^2+(p_3+p_2)^2\over m_\pi^2+
(p_3-p_2)^2} \bigg]  \,, \end{eqnarray}
with the $\tilde \mu$-derivative of the grand canonical potential due to static
$1\pi$-exchange and the NN-contact coupling:
\begin{eqnarray} \Omega_1'(\rho,T) &=&  {3 g_A^2 M \over 2f_\pi^2} 
\int_0^\infty \!\!{\dif} p_1\int_0^\infty \!\!{\dif} p_2\,\, d(p_1){d(p_2)\over
p_2} \bigg[ {(p_1+p_2)^3 \over m_\pi^2+(p_1+p_2)^2}\nonumber \\&& +{(p_1-p_2)^3
 \over m_\pi^2 +(p_1-p_2)^2} -2(\gamma_0+\gamma_1)p_1\bigg]\,, \end{eqnarray}
and the second $\tilde \mu$-derivative of the free nucleon gas part:
\begin{equation}\Omega_0'' (\rho,T)=- 4M\int_0^\infty  \!\! \dif p_1\,
\,{d(p_1) \over p_1} \,. \end{equation}
The first term in eq.(25) originates from taking into account the contributions
of (static) $1\pi$-exchange and the NN-contact coupling in the Legendre
transformation from the grand canonical potential to the free energy density
and from the perturbative shift of the chemical potential $\tilde \mu \to
\tilde \mu - \Omega_1'(\rho,T)/ \Omega_0''(\rho,T)$ (for details on that
procedure, see ref.\cite{kohn}). We have explicitly checked that the anomalous
contribution vanishes identically at $T=0$ for all densities $\rho$. This
vanishing, $\bar {\cal A}(\rho,0)=0$, is an automatic consequence of the 
Kohn-Luttinger-Ward theorem \cite{kohn} which of course hold in our case since 
the Fermi surface and the NN-interactions are invariant under spatial 
rotations. Furthermore, it is interesting to observe that the temperature and 
density dependent anomalous contribution $\bar {\cal A}(\rho,T)$ vanishes 
identically in the chiral limit $m_\pi = 0$. The latter feature has the
practical consequence that the behavior of nuclear matter at finite
temperatures $T\leq  30\,$MeV is only marginally influenced by the anomalous
contribution $\bar {\cal A}(\rho,T)$.

\begin{figure}
\begin{center}
\includegraphics[scale=0.5,clip]{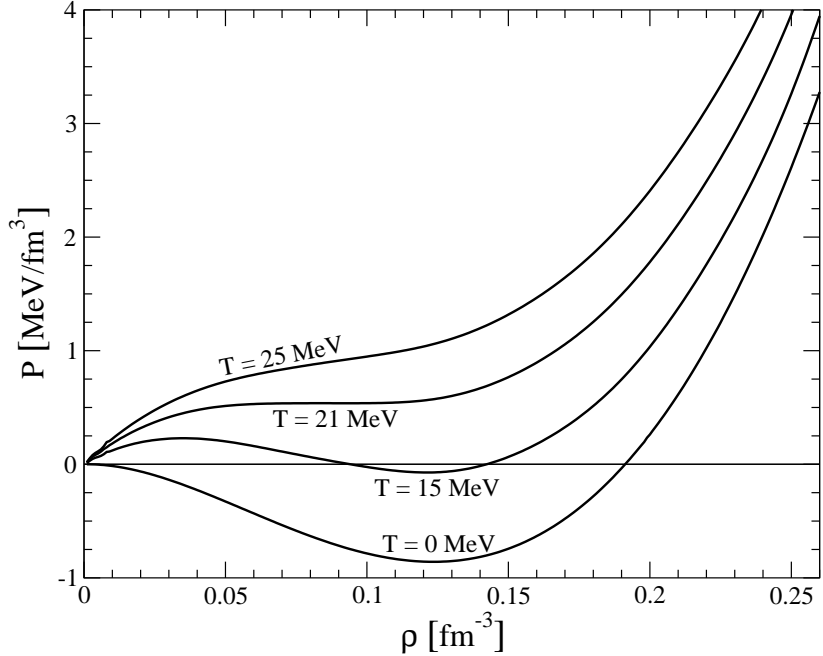}
\end{center}\vspace{-0.2cm}
\caption{Pressure isotherms $P(\rho,T)$ of isospin-symmetric nuclear matter for
the case of maximal saturation density. The coexistence region of the liquid
and gas phase ends at the critical point $\rho_c \simeq 0.09\fmd$ and $T_c
\simeq 21\,$MeV.}
\end{figure}

Fig.\,10 shows for the case of maximal saturation density ($\gamma_0=\gamma_1=
0.39)$ the calculated pressure isotherms $P(\rho,T) = \rho^2 \partial
\bar F(\rho,T)/\partial \rho$ of isospin-symmetric nuclear matter at four 
selected temperatures $T=0,\,15,\,21,\,25\,$MeV. As it should be these
curves display a first-order liquid-gas phase transition similar to that of the
van-der-Waals gas. The coexistence region between the liquid and the gas phase
(which has to be determined by the Maxwell construction) terminates at the
critical temperature $T_c$. From there on the pressure isotherms $P(\rho,T)$
grow monotonically with the nucleon density $\rho$. We find here a critical
temperature of $T_c \simeq 21\,$MeV. This is already compatible with the value
$T_c =(20\pm3)\,$MeV as extracted in ref.\cite{multifrag} from 
multi-fragmentation data in proton on gold collisions. Most other calculations
get still lower values of the critical temperature, typically $T_c \simeq 18\,
$MeV \cite{urbana,sauer,kapusta}. The reduction of $T_c$ in comparison to 
$T_c \simeq 25.5\,$MeV obtained in ref.\cite{liquidgas} from pion-exchange 
dynamics alone results from the less strong momentum dependence of the real 
single-particle potential $U(p,k_{f0})$ near the Fermi surface $p=k_{f0}$ (see
Fig.\,7). As a general rule $T_c$ grows with the effective nucleon mass
$M^*(k_{f0})$ at the Fermi surface. Because of that property $T_c \simeq 21\,
$MeV is the absolute minimum of the critical temperature possible within our 
perturbative calculation to fourth order.  

For the parameter set $(\gamma_0=0.39,\,\gamma_1=-0.75)$ constrained by the 
equation of state of pure neutron matter the unrealistic momentum dependence
of the associated real single-particle potential $U(p,k_{f0})$ (see Fig.\,6)
ruins the behavior of nuclear matter at finite temperatures. The critical
temperature of the liquid-gas phase transitions would lie in that case around 
$T_c \simeq 40\,$MeV, thus exceeding realistic values by more than a factor of
two.  

Evidently, the single-particle properties around the Fermi surface are decisive
for the thermal excitations and therefore they crucially influence the low 
temperature behavior of nuclear matter.

\section{Nuclear energy density functional}
The energy density functional is a general starting point for nuclear 
structure calculations within the framework of the self-consistent mean-field 
approximation \cite{reinhard}. In this context effective Skyrme forces 
\cite{sk3,sly,pearson} have gained much popularity because of their analytical 
simplicity and their ability to reproduce nuclear properties over the 
whole periodic table. In a recent work \cite{efun}, we have calculated the 
nuclear energy density functional which emerges from (leading and 
next-to-leading order) chiral pion-nucleon dynamics. The calculation in
ref.\cite{efun} included (only) the $1\pi$-exchange Fock diagram and the 
iterated $1\pi$-exchange Hartree and Fock diagrams. Therefore, the interest
here is on the additional contributions from the NN-contact interaction
appearing at the same order in the small momentum expansion.       

Going up to quadratic order in spatial gradients (i.e. deviations from
homogeneity) the energy density functional relevant for $N=Z$ even-even nuclei
reads \cite{efun}:  
\begin{equation} {\cal E}[\rho,\tau] = \rho\,\bar E(k_f)+\bigg[\tau-
{3\over 5} \rho k_f^2\bigg] \bigg[{1\over 2M}-{5k_f^2 \over 56 M^3}+F_\tau(k_f)
\bigg] + (\vec \nabla \rho)^2\, F_\nabla(k_f)\,,\end{equation}
with $\rho(\vec r \,) = 2k_f^3(\vec r\,)/3\pi^2$ the local nucleon density
(expressed in terms of a local Fermi momentum $k_f(\vec r\,)$) and $\tau
(\vec r\,)$ the local kinetic energy density. Note that we have left out in
eq.(28) the spin-orbit coupling term since the NN-contact vertex proportional 
to $\gamma_{0,1}$ does not generate any contribution to it. The starting point
for the construction of an explicit nuclear energy density functional ${\cal
E}[\rho,\tau]$ is the bilocal density matrix as given by a sum over the
occupied energy eigenfunctions: $\sum_{\alpha\in \rm occ}\Psi_\alpha( \vec r -
\vec a/2)\Psi_\alpha^\dagger(\vec r +\vec a/2)$. According to Negele and
Vautherin \cite{negele} it can be expanded in relative and center-of-mass
coordinate, $\vec a$ and $\vec r$, with expansion coefficients determined by
purely local quantities (nucleon density, kinetic energy density and spin-orbit
density). As outlined in sect.\,2 of ref.\cite{efun} the Fourier transform of 
the (so-expanded) density matrix defines in momentum space a medium insertion
for the inhomogeneous many-nucleon system:    
\begin{equation} \Gamma(\vec p,\vec q\,) =\int d^3 r \, e^{-i \vec q \cdot
\vec r}\,\theta(k_f-|\vec p\,|) \bigg\{1 +{35 \pi^2 \over 8k_f^7}(5\vec
p\,^2 -3k_f^2) \bigg[ \tau - {3\over 5} \rho k_f^2 - {1\over 4} \vec \nabla^2 
\rho \bigg] \bigg\}\,. \end{equation}
In comparison to eq.(5) in ref.\cite{efun} we have left out here the 
spin-dependent part (generating spin-orbit couplings) since it does not come 
into play. The strength function 
$F_\tau(k_f)$ in eq.(28) emerges via a perturbation on top of the density of
states $\theta(k_f-|\vec p\,|)$. As a consequence of that $F_\tau(k_f)$ can be
directly expressed in terms of the real single-particle potential $U(p,k_f)$ 
as:    
\begin{equation} F_\tau(k_f) = {35 \over 4k_f^7} \int_0^{k_f} dp\,
p^2(5p^2-3k_f^2)\, U(p,k_f) \,. \end{equation}
Note that any $p$-independent contribution, in particular, the cut-off 
dependent $\Gamma_{0,1}$-term in eq.(9) drops out. In the medium insertion 
eq.(29) $\tau-3\rho k_f^2/5$ is accompanied by $-\vec \nabla^2\rho/4$. 
After performing a partial integration one is lead to the following
decomposition:  
\begin{equation}F_\nabla(k_f) = {\pi^2 \over 8 k_f^2}\, {\partial F_\tau(k_f) 
\over  \partial k_f} +F_d(k_f) \,,\end{equation}
where $F_d(k_f)$ comprises all those contributions for which the $(\nabla\rho
)^2$-factor originates directly from the momentum-dependence of the 
interactions. We enumerate now the contributions to the strength functions 
$F_{\tau,d}(k_f)$ generated by the NN-contact vertex. From the second
pion-exchange diagram in Fig.\,1 with two medium insertions we get:
\begin{eqnarray} F_\tau(k_f) &=& {(\gamma_0+\gamma_1) g_A^4 M m_\pi^2 \over
(8\pi)^3 (u f_\pi)^4} \bigg\{\bigg({35u \over 2} -{63 \over 8u} -{5 \over 16 
u^3} \bigg) \ln(1+4u^2) \nonumber \\ && + {5\over 4u} -{117 u \over 2}-{29 u^3 
\over 3} +\bigg(4u^4+{175\over 4} \bigg) \arctan 2u \bigg\} \,, \end{eqnarray} 
\begin{equation} F_d(k_f) = {(\gamma_0+\gamma_1) g_A^4 M m_\pi\over
 4\pi (4 f_\pi)^4 (m_\pi^2+4k_f^2)} \,, \end{equation}
while the same diagram with three medium insertions gives:
\begin{eqnarray} F_\tau(k_f) &=& {35(\gamma_0+\gamma_1) g_A^4 M m_\pi^2 \over
(8\pi f_\pi)^4 u^7} \int_0^u{\dif}x\, x^2 \int_{-1}^1{\dif}y \bigg\{ 20u^3 xy [
s^2-\ln(1+s^2) ] \nonumber \\ && + \Big[2ux y +(u^2-x^2y^2) H \Big] \Big[
120 x y \arctan s +40 s x y(s^2-3) -15s^4 \nonumber \\ && + 
3(10+13u^2-20x^2-5x^2y^2)[s^2-\ln(1+s^2)]\Big]\bigg\} \,, \end{eqnarray}  
\begin{eqnarray} F_d(k_f) &=& -{(\gamma_0+\gamma_1) g_A^4 M  \over
\pi^2 m_\pi (4u f_\pi)^4} \int_0^u{\dif} x\, x^2 \int_{-1}^1 {\dif} y \nonumber
\\ && \times  \bigg\{{Hs^2 \over 8(1+s^2)^3}(15+10s^2+3s^4) +{u x y s^2(3+s^2)
\over 4(u^2-x^2y^2)(1+s^2)^2} \bigg\} \,. \end{eqnarray} 
In case of the last diagram in Fig.\,1 with three medium insertions we can even
solve the integrals and get:
\begin{equation} F_\tau(k_f) = {(\gamma^2_0+\gamma^2_1) g_A^4 M \over
(2\pi f_\pi)^4} \, {k_f^2 \over 108} (13-66 \ln2) \,.\end{equation} 
Evidently, there is no contribution to $F_d(k_f)$ from this last diagram since
the NN-contact vertex is momentum independent. The additional contributions to
$F_{\tau,d}(k_f)$ from $1\pi$-exchange and iterated  $1\pi$-exchange can be
found in sect.\,3 of ref.\cite{efun}. 

\bigskip

\begin{figure}
\begin{center}
\includegraphics[scale=0.5,clip]{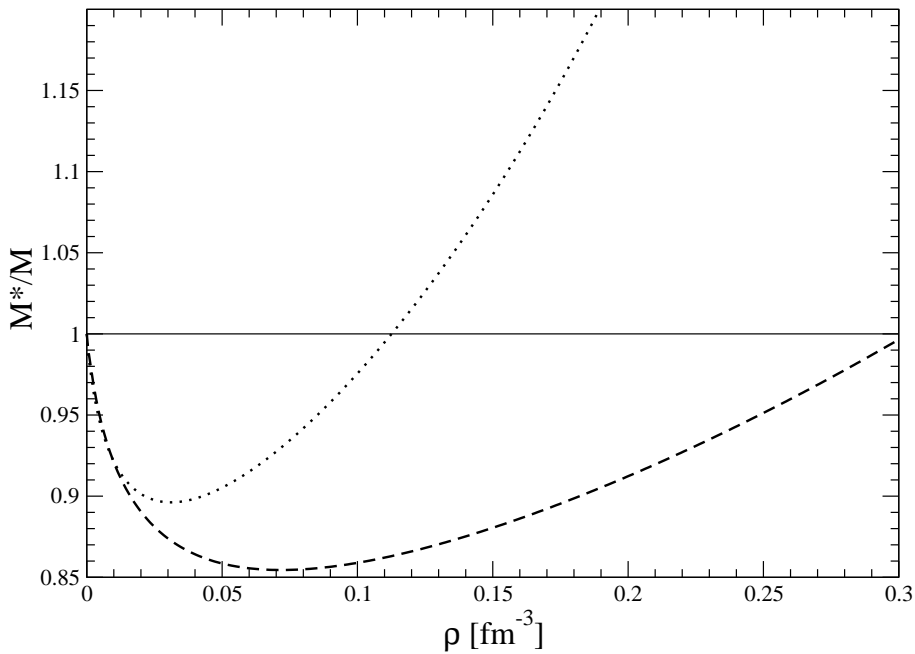}
\end{center}\vspace{-0.2cm}
\caption{The effective nucleon mass $\widetilde M^*(\rho)$ divided by the free
nucleon mass $M$ versus the nucleon density $\rho=2k_f^3/3\pi^2$. The dashed 
curve shows the case of maximal saturation density $\rho_0^{\rm max}=0.191
\fmd$. The dotted curve gives the result of the pion-exchange contributions
alone.}   
\end{figure}

\bigskip

The expression in eq.(28) multiplying the kinetic energy density $\tau(\vec
r\,)$ has the interpretation of a reciprocal density dependent effective
nucleon mass:  
\begin{equation} \widetilde M^*(\rho) = M \bigg[1-{5k_f^2 \over 28M^2}+ 2M\, 
F_\tau(k_f)\bigg]^{-1} \,. \end{equation}
In Fig.\,11, we show the ratio effective over free nucleon mass $\widetilde M^*
(\rho)/M$ as a function of the nucleon density $\rho=2k_f^3/3\pi^2$. The dashed
curve shows the case of maximal saturation density $\rho_0^{\rm max}=0.191
\fmd$ while the dotted curve gives the result of the pion-exchange 
contributions alone \cite{efun}. As a consequence of the more moderate momentum
dependence of the real single-particle potential $U(p,k_f)$ (see Fig.\,7) the
effective nucleon mass $\widetilde M^*(\rho)$ is now reduced for all densities
$\rho \leq 3\rho_0/2$. In comparison to most (non-relativistic) mean-field
calculations \cite{sk3,sly,pearson} the reduction by at most $15\%$ as obtained
here is a rather weak one. Note that the deviation of the effective nucleon 
mass $\widetilde M^*(\rho)$ from the free nucleon mass $M$ does not just grow 
linear with density, but there are very strong curvature effects. This has
mainly to do with the explicit presence of the small mass scale $m_\pi=135\,
$MeV in our calculation. For the parameter
set $(\gamma_0=0.39,\,\gamma_1=-0.75)$ constrained by the neutron matter 
equation of state the unrealistic momentum dependence of the corresponding 
real single-particle potential $U(p,k_{f0})$ (see Fig.\,6) would even lead 
to a singularity of $\widetilde M^*(\rho)$ at $\rho \simeq 0.16\fmd$. 

\bigskip

\begin{figure}
\begin{center}
\includegraphics[scale=0.5,clip]{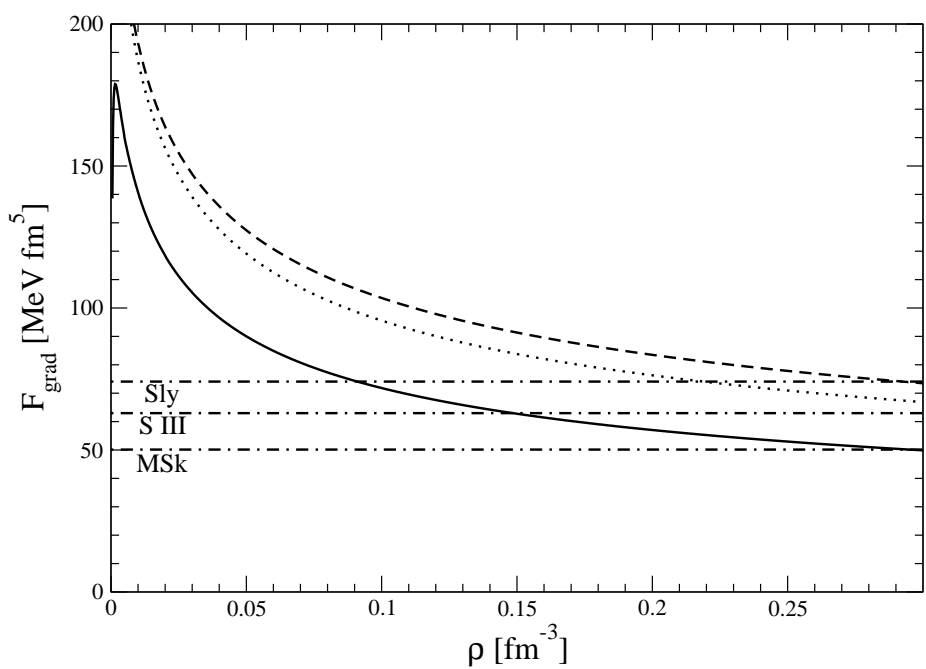}
\end{center}\vspace{-0.2cm}
\caption{The strength function $F_\nabla(k_f)$ related to the $(\vec \nabla 
\rho)^2$-term in the nuclear energy density functional versus the nucleon 
density $\rho=2k_f^3 /3\pi^2$. The full curve corresponds to the parameter set 
constrained by the neutron matter equation of state. The dashed curve shows
the case of maximal saturation density $\rho_0^{\rm max}=0.191\,\fmd$. The 
dotted curve gives the result of the pion-exchange contributions alone.} 
\end{figure}

\bigskip

In Fig.\,12, we show the strength function $F_\nabla(k_f)$ related to the 
$(\vec \nabla \rho)^2$-term in the nuclear energy density functional versus the
nucleon density $\rho=2k_f^3 /3\pi^2$. The full curve corresponds to the
parameter set constrained by the neutron matter equation of state. The dashed
curve shows the case of maximal saturation density $\rho_0^{\rm max}=0.191
\fmd$. For comparison, the dotted curve gives the result of the pion-exchange
contributions alone \cite{efun}. The three horizontal lines represent the 
constant values $F_\nabla(k_f)=[9t_1-(5+4x_2)t_2]/64$ of the Skyrme forces Sly
\cite{sly}, SIII \cite{sk3} and MSk \cite{pearson}. In the case of Sly and MSk 
we have performed averages over the various parameter sets Sly4-7 \cite{sly} 
and MSk1-6 \cite{pearson}. One observes that in the case of maximal saturation 
density (dashed line in Fig.\,12) the effects from the NN-contact interaction 
move the prediction for $F_\nabla(k_f)$ further away from the empirical values.
For the parameter set constrained by the neutron matter equation of state one 
gets from the NN-contact interaction a substantial lowering of the strength 
function  $F_\nabla(k_f)$ compared to the pure pion-exchange contributions
\cite{efun}. In the region $\rho_0/2 < \rho< 3\rho_0/2$ the prediction falls 
now into the empirically allowed band. The strong rise of $F_\nabla(k_f)$ at 
very low densities has to do with the presence of the small mass scale $m_\pi=
135\,$MeV, and with associated chiral singularities (of the form $m_\pi^{-2}$ 
and $m_\pi^{-1}$).    

Altogether, one finds a conflicting situation concerning the effects of the
NN-contact interaction on the strength functions $F_\tau(k_f)$ and 
$F_\nabla(k_f)$. The parameter set with an unrealistic $p$-dependence of 
single-particle potential $U(p,k_f)$, and therefore an unrealistic effective 
nucleon mass $\widetilde M^*(\rho)$, leads to an improved description of the 
strength function $F_\nabla(k_f)$ related to the $(\vec \nabla \rho)^2$-term in
the nuclear energy density functional. The inclusion of short-range NN-terms
does therefore not lead to a consistent improvement of inhomogeneous nuclear
matter properties.  
 
\section{Summary and conclusions}
In this work we have extended a recent chiral approach to nuclear matter by 
including the effects from the most general (momentum-independent) NN-contact 
interaction. By iterating the combination of one-pion exchange and this  
two-parameter contact-vertex to second order the emerging energy per particle 
exhausts all terms possible up-to-and-including fourth order in the small 
momentum expansion. Two cut-offs $\Lambda_{0,1}$, depending on the total
NN-isospin $I=0,1$, are introduced to regularize the (primitive) linear 
divergences $\int_0^\infty {\dif}l\,1$ of some three-loop in-medium diagrams. 
Effectively, the cut-offs $\Lambda_{0,1}$ parameterize the strength of a term
linear in density, see eqs.(2,5). The equation of state of pure neutron matter,
$\bar E_n(k_n)$, can be reproduced very well up to quite high neutron densities
of $\rho_n=0.5\fmd$ by adjusting the strength $\gamma_1$ of a repulsive 
$nn$-contact interaction. With $\gamma_1=-0.75$ it is comparably strong as the
$1\pi$-exchange. 
 
Binding and saturation of isospin-symmetric nuclear matter is a generic feature
of the present perturbative calculation. Fixing the maximum binding energy per 
particle to $-\bar E(k_{f0}) \\= 15.3\,$MeV we find that any possible 
equilibrium density $\rho_0$ lies below $\rho_0^{\rm max}=0.191\fmd$. The
constraint $\gamma_1=-0.75$ from the neutron matter equation of state leads, 
however, to a somewhat too low saturation density of $\rho_0 =0.134 \fmd$. 
Fixing on the other hand the saturation density to $\rho_0=0.158\fmd$ 
\cite{sick} we find that the maximum binding energy per particle $-\bar 
E(k_{f0})$ lies above $10.7\,$MeV.  

We have also investigated the effects of the NN-contact interaction on the
complex single-particle potential $U(p,k_f)+i\,W(p,k_f)$. We have found that
the effective  nucleon mass at the Fermi-surface is bounded from below by 
$M^*(k_{f0}) \geq 1.4M$. The minimal effective nucleon mass occurs
simultaneously together with the maximal saturation density. The corresponding 
parameter set $\gamma_0 =\gamma_1=0.39$ describes the situation in which the
explicitly introduced short-range NN-terms cancel almost exactly the contact 
pieces of $1\pi$-exchange. The parameter set constrained by the neutron matter 
equation of state leads to unrealistic real and imaginary single-particle 
potentials, $U(p,k_{f0})$ and $W(p,k_{f0})$, as in the scheme of Lutz et al. 
\cite{lutz,lutzcontra}. The downward bending of the asymmetry energy $A(k_f)$ 
above nuclear matter saturation density is a generic feature of our 
perturbative (fourth-order) calculation. The value of the asymmetry energy at
saturation density $A(k_{f0})$ is rather stable and comes out close to its
empirical value $A(k_{f0})\simeq 33\,$MeV.   

The properties of nuclear matter at finite temperatures have also been 
investigated. The critical temperature of the liquid-gas phase transition has 
a minimal value of $T_c \simeq 21\,$MeV. Good single-particle properties around
the Fermi surface are a necessary condition for a realistic behavior of 
isospin-symmetric nuclear matter at finite temperatures $T<40$\,MeV. 

The influence of the NN-contact interaction on the effective nucleon mass 
$\widetilde M^*(\rho)$ and the strength function $F_\nabla(k_f)$ of the 
$(\vec \nabla\rho)^2$-term in the nuclear energy density functional 
${\cal E}[\rho,\tau]$ have also been studied. For the parameter set with 
reduced $\widetilde M^*(\rho)$ the prediction for $F_\nabla(k_f)$ is moved
further away from the empirical band. On the other hand the parameter set with
unrealistic  $\widetilde M^*(\rho)$ substantially improves $F_\nabla(k_f)$.

Altogether, there is within this complete fourth order calculation no "magic" 
set of short-range parameters with which one could reproduce simultaneously 
and  accurately all semi-empirical properties of nuclear matter. The 
conditions for a good neutron matter equation of state and for good
single-particle properties (and consequently a realistic finite temperature
behavior) are in fact mutually exclusive. This clearly points at inherent 
limitations of the perturbative chiral approach truncated at fourth order in 
the small momentum expansion. While complete calculation at fifth order or even
sixth order are rather demanding (four-loop and five-loop diagrams need to be
evaluated) one might also combine the chiral approach (to nuclear matter) with 
non-perturbative methods as it is done commonly for free NN-scattering
\cite{epel,mach}.   

\section*{Appendix}
In this appendix we take the opportunity to write down closed form analytical 
expressions for the three-body and four body contributions $W_{3,4}(p,k_f)$ to 
the imaginary single-particle potential as they arise from the iterated 
$1\pi$-exchange Hartree diagram. Solving the integrals in eq.(22) of
ref.\cite{pot} gives for the three-body term:
\begin{eqnarray} W_3(p,k_f) &=& {g_A^4Mm_\pi^4\over(8\pi)^3 f_\pi^4}\bigg\{
35+151u^2+25u^4-{197x^2 \over 3} -2u^2x^2-7x^4 +18\ln(1+4x^2) \nonumber \\ & & 
+\bigg({21\over 2x}-30x\bigg)\arctan2x +\Big[\arctan(u+x)- \arctan(u-x) \Big] 
\nonumber \\ & & \times {1\over 2x}\Big[(x^2-u^2)^3 +9(4u^2+15) (x^2-u^2)-56
\Big]  \nonumber \\ & & +4\Big( 3x^2-9-{2u^3\over x}-9u^2\Big)\ln[1+(u+x)^2] 
 \nonumber \\ & & + 4\Big(3x^2-9+{2u^3\over x}-9u^2 \Big) \ln[1+(u-x)^2] 
\bigg\}\,, \end{eqnarray}
with $x=p/m_\pi$ and $u = k_f/m_\pi$. After complete evaluation of its integral
representation given in eq.(24) of ref.\cite{pot} the four-body Hartree term 
reads: 
\begin{eqnarray} W_4(p,k_f) &=&{g_A^4Mm_\pi^4\over (8\pi)^3f_\pi^4}\bigg\{
{63 x^4 \over 5}-13u^4-7 -2u^2(49+3x^2) +{142x^2 \over 3}-36\ln(1+4x^2) 
\nonumber \\ & & +\bigg(60x-{21\over x}\bigg)\arctan2x +\Big[\arctan(u+x)- 
\arctan(u-x) \Big] \nonumber \\ & & \times {1\over 2x}\Big[(u^2-x^2)^3-9x^4
-18u^2x^2+27u^4-135x^2+75u^2+49\Big]  \nonumber \\ & &+4(9+6u^2-2x^2)
\ln\Big([1+(u+x)^2][1+(u-x)^2]\Big) \bigg\}\,.\end{eqnarray}
Combining these novel results with the two-body Hartree term $W_2(p,k_f)$
written in eq.(20) of ref.\cite{pot} one immediately verifies Luttinger theorem
\cite{luttinger}:
\begin{equation} W_2(k_f,k_f)+W_3(k_f,k_f)+ W_4(k_f,k_f)=0 \,, \end{equation}
and the zero sum-rule \cite{pot}:
\begin{equation} {3\over k_f^3}\int_0^{k_f} dp\, p^2 \bigg[{1\over 2}W_2(p,k_f)
+{1\over 3}W_3(p,k_f)+ {1\over 4}W_4(p,k_f)\bigg] = 0\,. \end{equation} 
The latter statement expresses the fact that the energy per particle $\bar
E(k_f)$ can be reconstructed from the two-, three- and four-body contributions
to the single-particle potential (graphically this is done by closing the open
nucleon line of self-energy diagrams) and that no imaginary part is left over 
from this procedure.     

\end{document}